\documentclass[aps,prl,preprint,tightenlines,superscriptaddress,showpacs,byrevtex]{revtex4}

\usepackage{graphicx} 
\usepackage{dcolumn}  
\usepackage{amsmath, amsthm, amssymb} 

\newcommand{\ra}{\rightarrow}
\newcommand{\lam}{\Lambda}
\newcommand{\ks}{K_S}
\newcommand{\lamst}{\Lambda(1520)}
\newcommand{\tht}{\Theta^+}
\newcommand{\thst}{\Theta^{\ast ++}}
\newcommand{\thc}{\Theta_c^0}
\newcommand{\thcst}{\Theta_c^{\ast +}}
\newcommand{\mev}{\mathrm{MeV}/c^2}
\newcommand{\gev}{\mathrm{GeV}/c^2}
\newcommand{\mevc}{\mathrm{MeV}/c}
\newcommand{\gevc}{\mathrm{GeV}/c}
\newcommand{\cm}{\mathrm{cm}}

\newcommand{\km}{K^-}
\newcommand{\kp}{K^+}
\newcommand{\de}{\Delta E}
\newcommand{\mbc}{M_\mathrm{bc}}

\begin{document}

\vbox to 16mm{
                 \vss\hbox{\resizebox{!}{3cm}{
                 \includegraphics{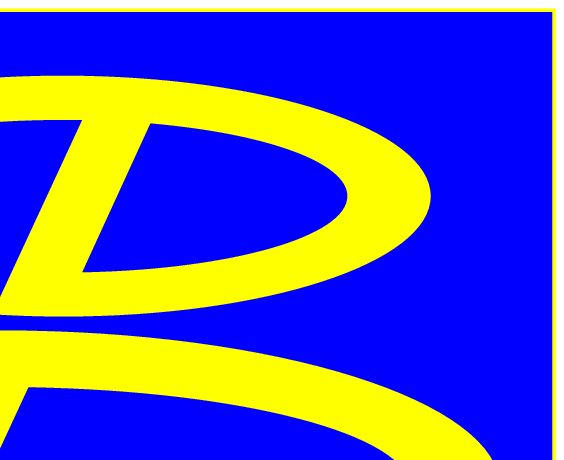}}}}\vspace{-1cm}

\title{\quad\\[0.5cm] \boldmath
  Search for pentaquarks at Belle
}

\author{R.~Mizuk}

\affiliation{Institute for Theoretical and Experimental Physics,
B.Cheremushkinskaya, 25,
117259 Moscow, Russia \\
E-mail: mizuk@itep.ru \\
(for the Belle Collaboration)}


\tighten

\begin{abstract}
We search for the strange pentaquark $\tht$
using kaon interactions in the material of the Belle detector. 
No signal is observed in the $p\ks$ final state, 
while in the $pK^-$ final state 
we observe $\sim 1.6 \cdot 10^4$ $\lamst \to p K^-$ decays. 
We set an upper limit on the ratio of $\tht$ to $\lamst$ yields
$\sigma(\tht)/\sigma(\lamst) < 2\%$ at 90\% CL, assuming 
that the $\tht$ is narrow. 
We also report on searches for strange and charmed pentaquarks 
in $B$ meson decays. 
These
results are obtained from a $155\,{\rm fb}^{-1}$ data sample collected
with the Belle detector near the $\Upsilon(4S)$ resonance, at the KEKB
asymmetric energy $e^+ e^-$ collider.
\end{abstract}

\pacs{13.25.Hw, 13.75.Jz, 14.20.Jn, 14.20.-c}

\maketitle

{\renewcommand{\thefootnote}{\fnsymbol{footnote}}}

\section{Introduction}

Until recently, all reported particles could be understood as
bound states of three quarks or a quark and an antiquark. 
QCD predicts also more complicated configurations such as glueballs $gg$,
molecules $q\bar{q}\,q\bar{q}$ and pentaquarks $qqq\,q\bar{q}$. 
Recently, observations of the pentaquark $\tht =uudd\bar{s}$ have been
reported in the decay channels $K^+n$~\cite{leps} and
$p\ks$~\cite{diana}. 
Many experimental groups have confirmed this observation and 
the isospin 3/2 members of the same pentaquark multiplet 
have also been observed~\cite{na49}.
Evidence for the charmed pentaquark $\thc =uudd\bar{c}$
has also been seen~\cite{h1}. 
The topic attracts enormous theoretical interest. 
However the existence and properties of pentaquarks remain a mystery. 
Some experimental groups do not see the pentaquark signals. 
The non-observing experiments correspond to higher
center-of-mass energies. It has been argued~\cite{Titov}
that pentaquark production is suppressed in the fragmentation regime
at high energies.

Charged and neutral kaons are copiously produced at Belle. 
We treat kaons as projectiles and the detector material as a target,
and search for strange pentaquark 
\emph{formation}, $K N \to \tht$,
and \emph{production}, $K N \to \tht X$. 
The kaon momentum spectrum is soft, with a most probable momentum of
only $0.6\,\gevc$. Therefore we can search for $\tht$ formation
in the low energy region.

We also search for strange and charmed pentaquarks 
in the decays of $B$ mesons,
where the suppression of production observed in $s$ channel
$e^+e^-$ collisions~\cite{babar} may be absent. Studies of $B$ meson decays have proved
to be very useful for discoveries of new particles 
(such as P-wave $c\bar{q}$ states), 
therefore it is interesting to search for
pentaquarks in $B$ decays although no firm theoretical predictions for branching
fractions exist.

\section{Detector and data set}

These studies are performed using a data sample of $140
\,\mathrm{fb}^{-1}$ collected at the $\Upsilon(4S)$ resonance and
$15\,\mathrm{fb}^{-1}$ at an energy $60\,{\mathrm{MeV}}$ below the
resonance. The data were collected with the Belle
detector~\cite{BELLE_DETECTOR} at the KEKB asymmetric energy $e^+ e^-$
storage rings~\cite{KEKB}.

The Belle detector is a large-solid-angle magnetic spectrometer that
consists of a three layer silicon vertex detector (SVD), a 50-layer
cylindrical drift chamber (CDC), a mosaic of aerogel threshold
Cherenkov counters (ACC), a barrel-like array of time-of-flight
scintillator counters (TOF), and an array of CsI(Tl) crystals (ECL)
located inside a superconducting solenoidal coil that produces a 1.5~T
magnetic field. An iron flux return located outside the coil is
instrumented to detect muons and $K_L$ mesons (KLM). 

The proton, kaon and charged pion candidates are identified 
based on the $dE/dx$, TOF and Cherenkov light yield information 
for each track. 
$\ks$ candidates are reconstructed via the $\pi^+ \pi^-$ decays 
and must have an invariant mass consistent with the nominal $\ks$
mass. The $\ks$ candidate is further required to have 
a displaced vertex and a momentum direction consistent 
with the direction from its production to decay vertices.  

\section{Search for $\tht$ using kaon interactions
in the detector material.}

\begin{figure}[t]
\centering
\begin{picture}(550,220)
\put(3,-10){\includegraphics[width=0.95\textwidth]{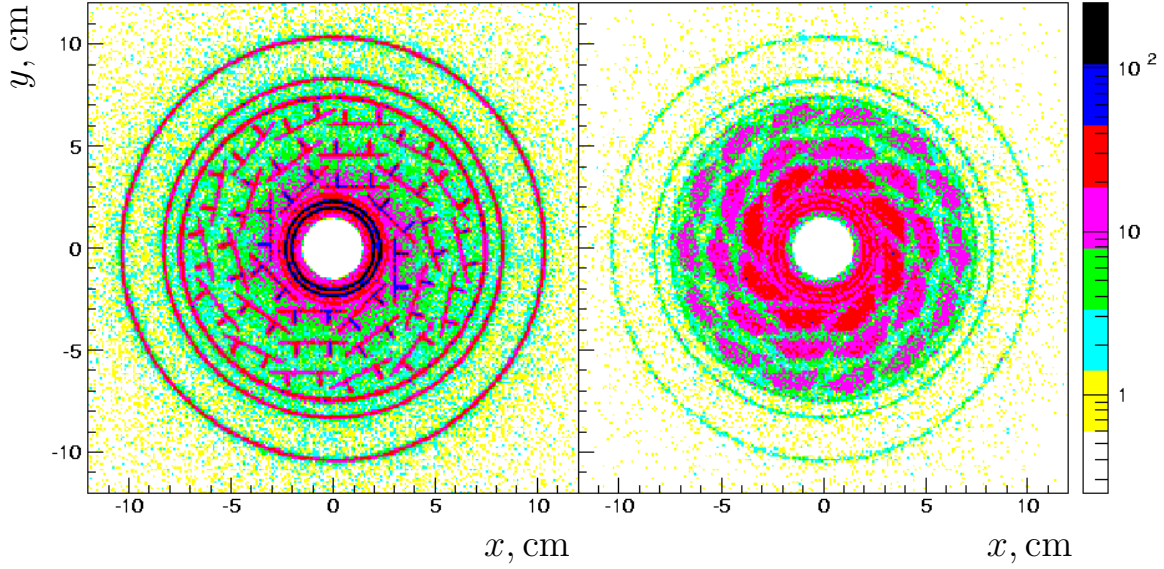}}
\put(11,180){\rotatebox{90}{\large $y,\cm$}}
\put(190,3){\large $x, \cm$} 
\put(380,3){\large $x, \cm$} 
\end{picture}
\caption{The $xy$ distribution of secondary $pK^-$ vertices 
for the barrel (left) and endcap (right) parts of the detector.}  
\label{xy}
\end{figure}

The analysis is performed by selecting $pK^-$, $pK^+$ and $p\ks$
secondary vertices.  
The protons and kaons are required not to originate 
from the region around the run-averaged interaction point (IP). 
The proton and kaon candidate are combined and the $pK$ vertex is
fitted. 
The $xy$ distribution of the secondary $pK^-$ vertices is shown in 
Fig.~\ref{xy} for the barrel part (left) and for the endcap part
(right) of the detector. 
The double wall beam pipe, three layers of SVD, the SVD cover and the two
support cylinders of the CDC are clearly visible. The $xy$ distributions
for secondary $pK^+$ and $p\ks$ vertices are similar. 

\begin{figure}[tbh]
\centering
\begin{picture}(550,190)
\put(20,115){\rotatebox{90}{N/5$\mev$}} 
\put(230,115){\rotatebox{90}{N/5$\mev$}} 
\put(125,5){$M(pK^+),\gev$} 
\put(305,5){$M(pK^-,p\ks),\gev$} 
\put(20,-10){\includegraphics[width=8cm]{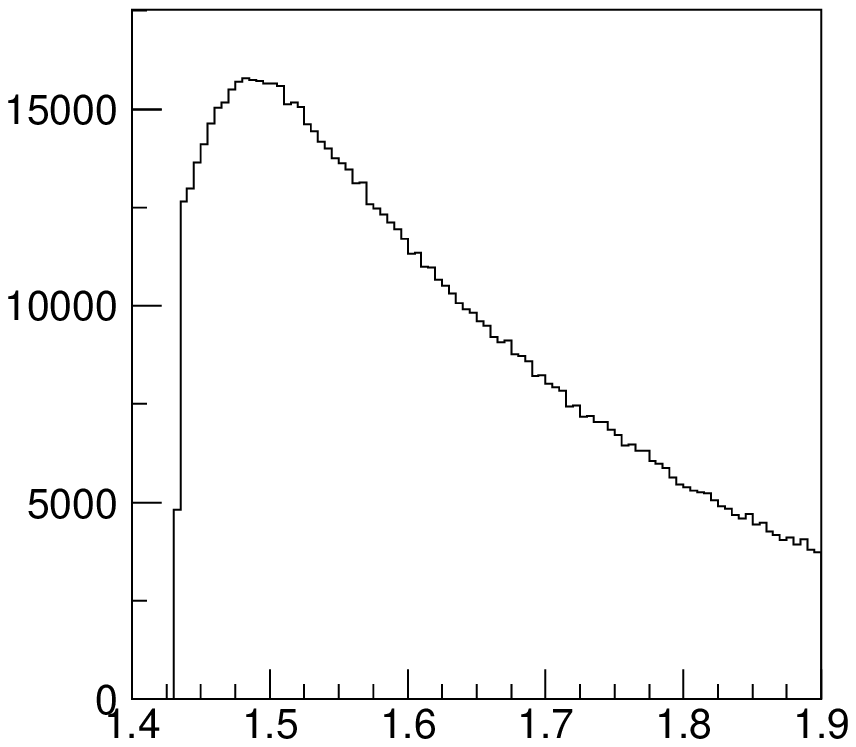}}
\put(230,-10){\includegraphics[width=8cm]{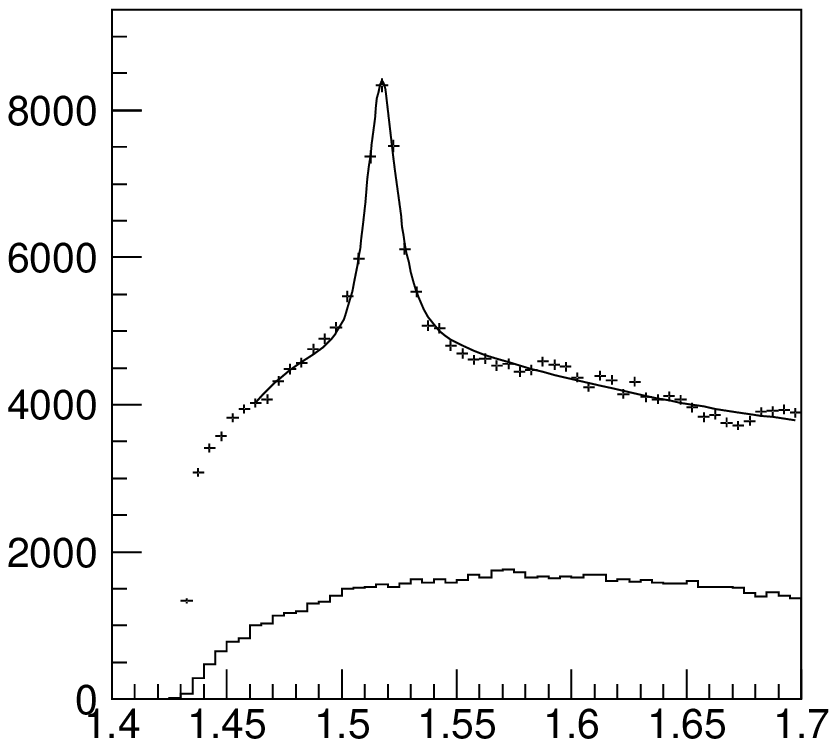}}
\end{picture}
\caption{Mass spectra of $pK^+$ (left), $pK^-$ (right, points with
  error bars) and $p\ks$ (right, histogram) secondary pairs. The fit
  is described in the text.} 
\label{m_pk}
\end{figure}

The mass spectra for $pK^-$, $pK^+$ and $p\ks$ secondary vertices
are shown in Fig.~\ref{m_pk}. No significant structures are observed
in the $M(pK^+)$ or $M(p\ks)$ spectra, while in the $M(pK^-)$ spectrum a
$\lamst$ signal is clearly visible. 
We fit the $pK^-$ mass spectrum to a sum of a $\lamst$
probability density function (p.d.f.) and a threshold function. 
The signal p.d.f.\ is a D-wave Breit-Wigner shape 
convolved with a detector resolution function
($\sigma \sim 2\,\mev$). 
The $\lamst$ parameters obtained from the
fit are consistent with the PDG values~\cite{pdg}.
The $\lamst$ yield, defined as the signal p.d.f.\ integral over 
the $1.48$--$1.56\,\gev$ mass interval ($2.5\Gamma$), is $15519\pm 412$
events. 

The $p\ks$ mass spectrum is fitted to a sum of a $\tht$
signal p.d.f. and a third order polynomial. 
The $\tht$ signal shape can be rather
complicated because of possible rescattering of particles inside
nuclei~\cite{sibirtsev}.
In order to compare our result with other experiments we assume that
the signal is narrow and its shape is 
determined by the detector resolution ($\sim2\,\mev$). 
For $m=1540\,\mev$ the fit result is $29\pm 65$
events. Using the Feldman-Cousins method
of upper limit evaluation~\cite{feldman-cousins} we obtain
$N<94$~events at the 90\% CL.
We set an upper limit on the ratio of $\tht$ to $\lamst$ yields
corrected for the efficiency and branching fractions:
\[
\frac{N_{\rm observed}(\tht)}{N_{\rm observed}(\lamst)}
\frac{\epsilon(pK^-)}{\epsilon(p\ks)}
\frac{\mathcal{B}(\lamst\ra pK^-)}{\mathcal{B}(\tht\ra p\ks)\cdot
  \mathcal{B}(\ks\ra\pi^+\pi^-)}<2\%~(90\%~\text{CL}).
\]
It is assumed that $\mathcal{B}(\tht\ra p\ks)=25\%$. 
We take $\mathcal{B}(\lamst\ra pK^-) = 
\frac{1}{2} \mathcal{B}(\lamst\ra N\bar{K}) =
\frac{1}{2}(45\pm 1)\%$~\cite{pdg}.
The ratio of efficiencies for 
$\tht \ra p\ks$ and 
$\lamst \ra p\km$ 
of $37\%$ is obtained from the Monte Carlo (MC) simulation. 
Our limit is much smaller than the results reported by many
experiments which observe $\tht$. 
For example it is two orders of magnitude smaller than the value
reported by the \mbox{HERMES} Collaboration~\cite{HERMES}.
We do not know any physical explanation for such a difference.

\begin{figure}[tbh]
\centering
\begin{picture}(550,190)
\put(20,110){\rotatebox{90}{N$/0.2\,\gev$}} 
\put(235,110){\rotatebox{90}{N$/50\,\mev$}}
\put(115,5){$p(\lamst),\,\gev$} 
\put(340,5){$p(K^\pm),\,\gev$} 
\put(20,-10){\includegraphics[width=8cm]{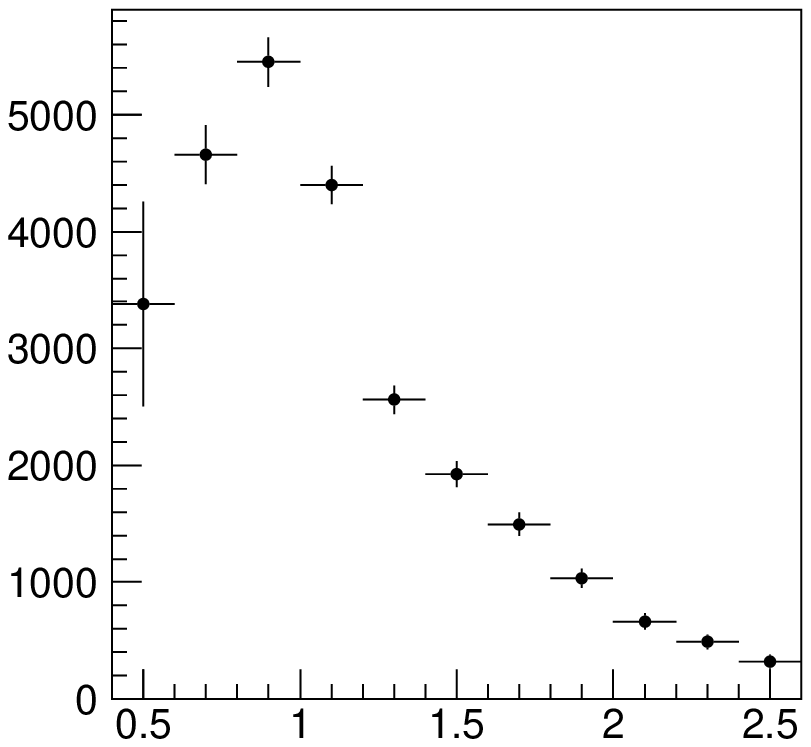}}
\put(230,-10){\includegraphics[width=8cm]{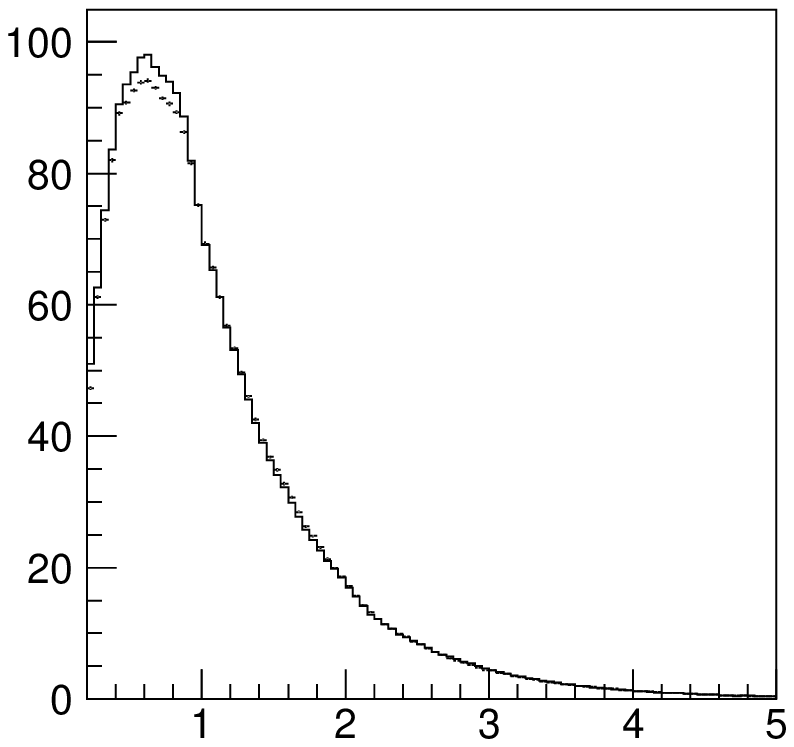}}
\end{picture}
\caption{Left: momentum spectrum of the $\lamst$.
  Right: momentum spectrum of $\km$ (points with error bars)
  and $\kp$ (solid histogram). }
\label{mom}
\end{figure}

The momentum spectrum of the produced $\lamst$ is shown in
Fig.~\ref{mom}~(left). 
This spectrum is obtained from fitting $M(pK^-)$ in momentum
bins and correcting for the efficiency obtained from MC. 
The $K^-$ should have a $440\,\mevc$ momentum to produce $\lamst$ on a 
proton at rest. Even in the presence of Fermi motion with a typical
momentum of $150\,\mevc$, $\lamst$ produced in the {\it formation} channel
should be contained in the first momentum bin, 
0.4 to $0.6\,\gev$. Therefore most of the $\lamst$ are produced in the
{\it production} channel. The projectiles that can produce $\lamst$
are $K^-$, $\ks$, $K_L$, $\lam$. The momentum spectra of $K^-$ and
$K^+$ are given in Fig.~\ref{mom}~(right). The spectra are
corrected for efficiency and for contamination from other
particle species. 
It is not likely that $\lamst$ production is dominated by
interactions induced by $\lam$ projectiles, 
because the $\lamst$ momentum spectrum is too soft. 
Even at the threshold of the $\lam N\to\lamst p$ reaction 
the $\lamst$ momentum is $\sim 1.1\,\gevc$. 

\begin{figure}[tbh]
\centering
\begin{picture}(550,190)
\put(115,135){\rotatebox{90}{N/0.5$\cm$}} 
\put(240,5){$Distance,\cm$} 
\put(115,-10){\includegraphics[width=8cm]{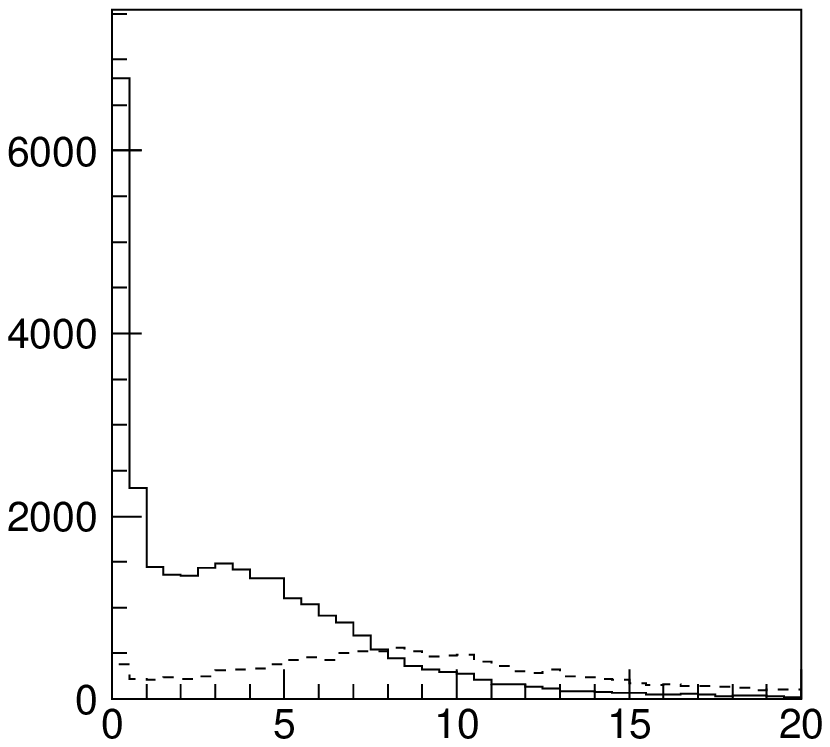}}
\end{picture}
\caption{Distance from $p\km$ secondary vertex to the nearest track
  (solid histogram) and to the nearest $K^+$ (dashed histogram).}
\label{kp_tag}
\end{figure}

To demonstrate that non-strange particles do not produce $\lamst$ we
study the 
$pK^-$ vertices accompanied by a $K^+$ tag. The distance from the
$pK^-$ vertex to the nearest $K^+$ is plotted in Fig.~\ref{kp_tag} 
as a dashed histogram. For comparison the distance to any track is
plotted as a solid histogram. 
The peak at zero corresponds to the vertices with additional
tracks. 
The much smaller peak at zero for the $\kp$ tagged vertices 
leads us to the conclusion that most $\lamst$ are produced by strange
projectiles. 

\section{Search for pentaquarks in $B$ meson decays}

In this analysis we search for $\tht$ and $\thst$ 
(an isovector pentaquark predicted in some models~\cite{bkm}) 
in the decays 
$B^0 \to \tht \bar{p}$ followed by $\tht \to p K_S$, and 
$B^+ \to \thst \bar{p}$ followed by $\thst \to p K^+$, respectively 
(inclusion of charge conjugated modes is implied throughout this section). 
We also search for $\thc$ in the decay
$B^0 \to \thc \bar{p} \pi^+$ followed by $\thc \to D^{(\ast)-} p$,
and $\thcst$ (the charmed analogue of $\thst$) in the decay
$B^0 \to \thcst \bar{p}$ followed by $\thcst \to \bar{D^0} p$.
We reconstruct $D$ mesons in the decay modes 
$D^{\ast +}\to D^0\pi^+$, $D^0\to K^-\pi^+$ and $D^-\to K^-\pi^+\pi^+$. 
The dominant background arises from the continuum 
$e^+ e^- \to q \bar q$ process. 
It is suppressed using event shape variables 
(the continuum events are jet-like, 
while the $B\bar B$ events are spherically symmetric).

\begin{figure}[ht]
\begin{center}
\resizebox{12cm}{!}{\includegraphics{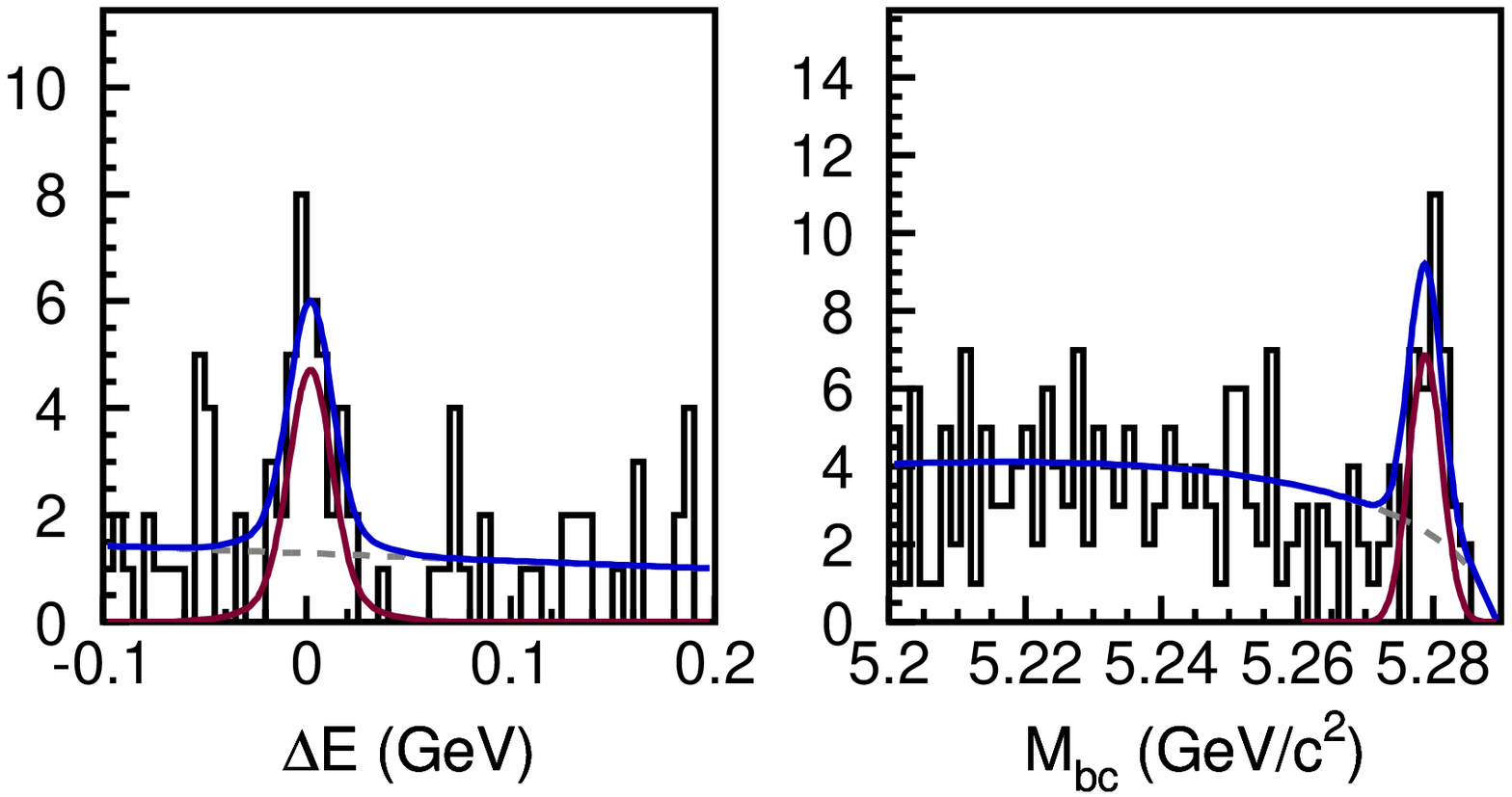}}
\resizebox{12cm}{!}{\includegraphics{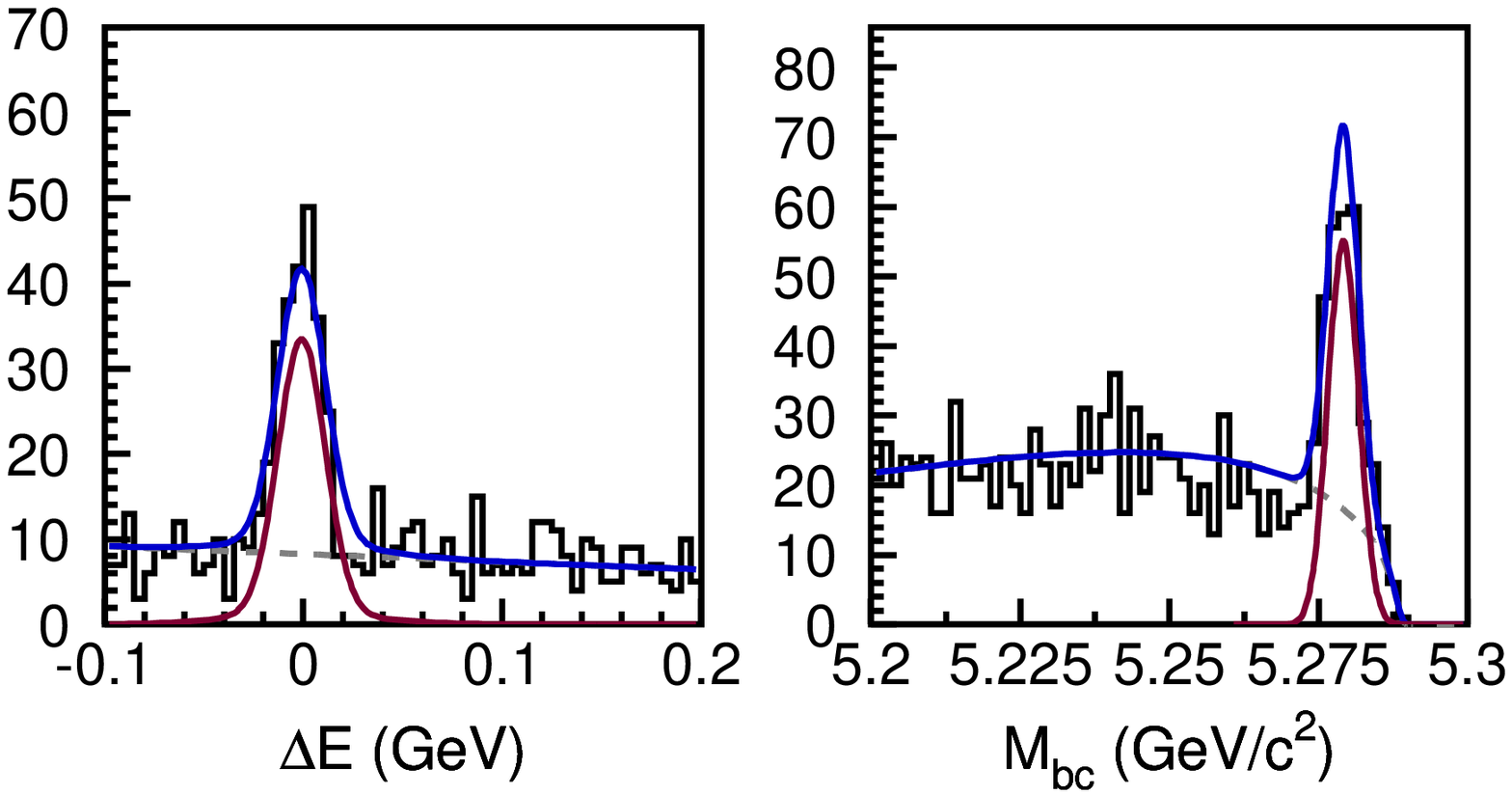}}
\caption{ $\Delta E$ and $M_\mathrm{bc}$ distributions for 
  $p \bar{p} K_S$ (top), and $p \bar{p} K^+$ (bottom) modes. 
The curves represent the fit results. 
}
\label{ppbark}
\end{center}
\end{figure}

The $B$ decays are identified by their CM energy difference, 
\mbox{$\de=(\sum_iE_i)-E_{\rm beam}$}, and the beam constrained mass,
$\mbc=\sqrt{E_{\rm beam}^2-(\sum_i\vec{p}_i)^2}$, where $E_{\rm beam}$
is the beam energy and $\vec{p}_i$ and $E_i$ are the momenta and
energies of the decay products of the $B$ meson in the CM frame.
The $\de$ distribution (with $\mbc>5.27\,\gev$)
and  $\mbc$ distribution (with $|\de|< 0.05\,\gev$) 
for the $B^0 \to p \bar{p} K_S$ and $B^+ \to p \bar{p} K^+$ decays 
are shown in Fig.~\ref{ppbark}. 
The signal yields are extracted by performing unbinned 
maximum likelihood fits to the sum of signal and background distributions 
in the two dimensional ($\mbc$, $\de$) space. The signal 
distributions are determined from MC, whereas the background 
distributions are determined from the $\de$ and $\mbc$ 
sideband data samples. 
The fits  give $28.6^{+6.5}_{-5.8}$ and 
$216.5 ^{+17.3}_{-16.6}$ signal yields for the $p \bar{p} K_S$ and 
$p \bar{p} K^+$ modes, respectively. 
For the region 
$1.53~\mathrm{GeV}/c^2 < M_{p K_S} < 1.55~\mathrm{GeV}/c^2$, corresponding 
to the reported $\tht$ mass, we find no signal.
Since there is only a theoretical conjecture for the $\Theta^{*++}$, we check the 
$1.6~\mathrm{GeV}/c^2 < M_{p K^+} < 1.8~\mathrm{GeV}/c^2$ region and find 
no signal. Assuming both states are narrow, we set the upper limits
\begin{align*}
\frac{{\mathcal B}(B^0 \to \Theta^+ \bar{p}) \times {\mathcal B}(\Theta^+ \to p K_S)}
     {{\mathcal B}(B^0 \to p \bar{p} K_S)} & < 22\%~(90\%~\text{CL})
\;\text{and}\\[1.5ex]
\frac{{\mathcal B}(B^+ \to \Theta^{*++}\bar{p}) \times 
              {\mathcal B}(\Theta^{*++} \to p K^+)}
     {{\mathcal B}(B^+ \to p \bar{p} K^+)} & < 2\%~(90\%~\text{CL}).
\end{align*}

\begin{figure}[ht]
\begin{center}
\resizebox{5.4cm}{!}{\includegraphics{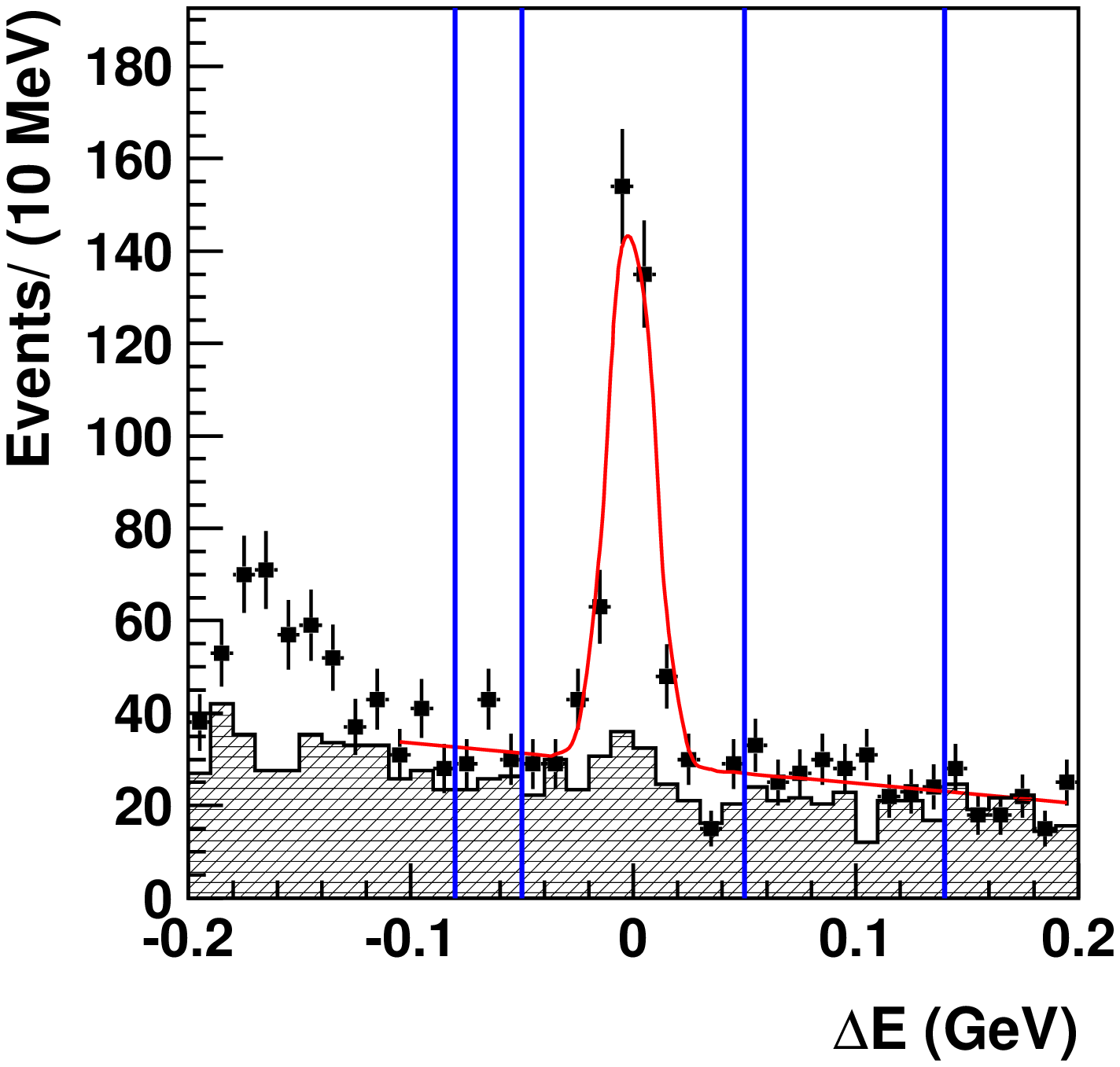}}
\resizebox{5.4cm}{!}{\includegraphics{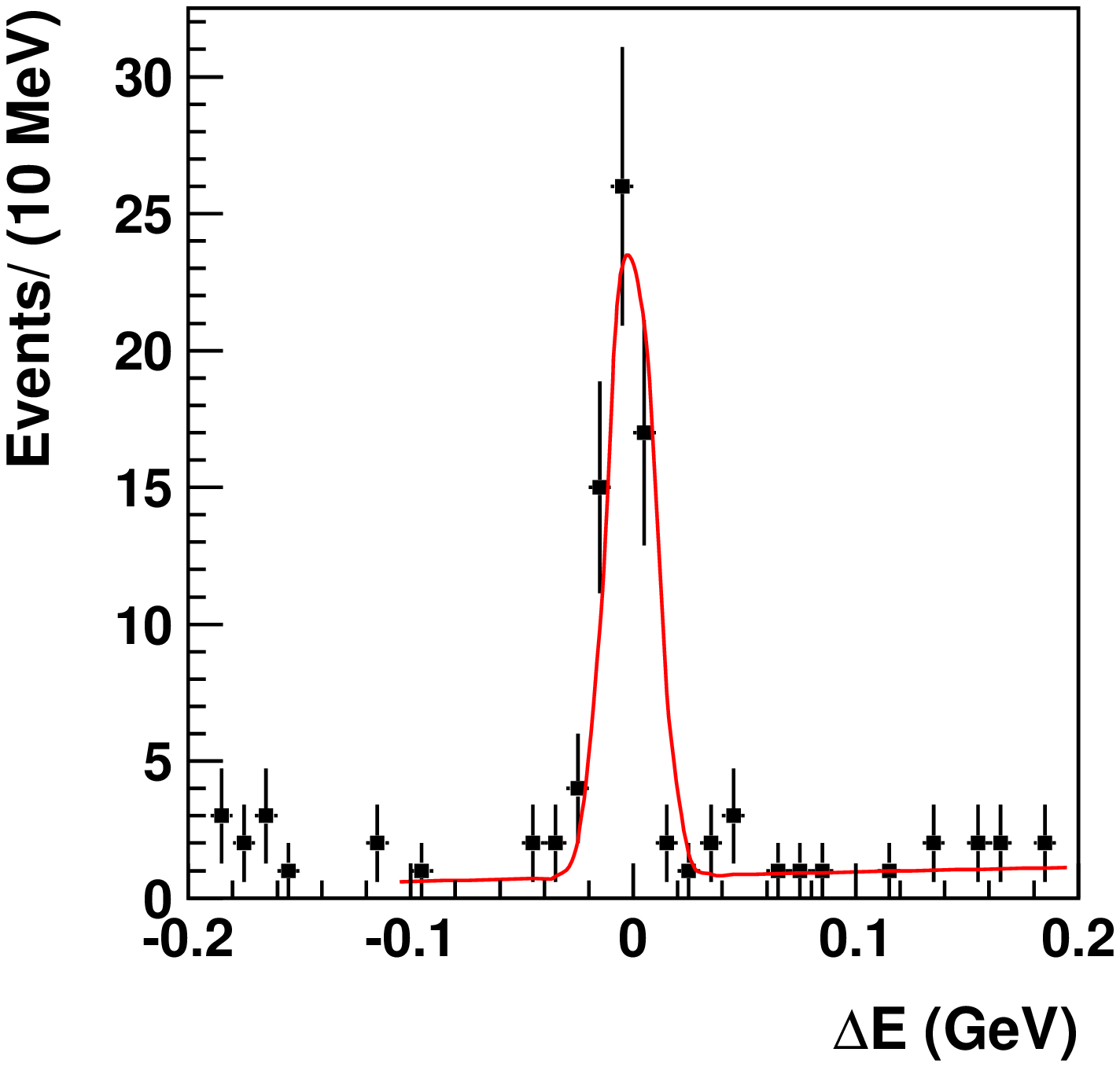}}
\resizebox{5.4cm}{!}{\includegraphics{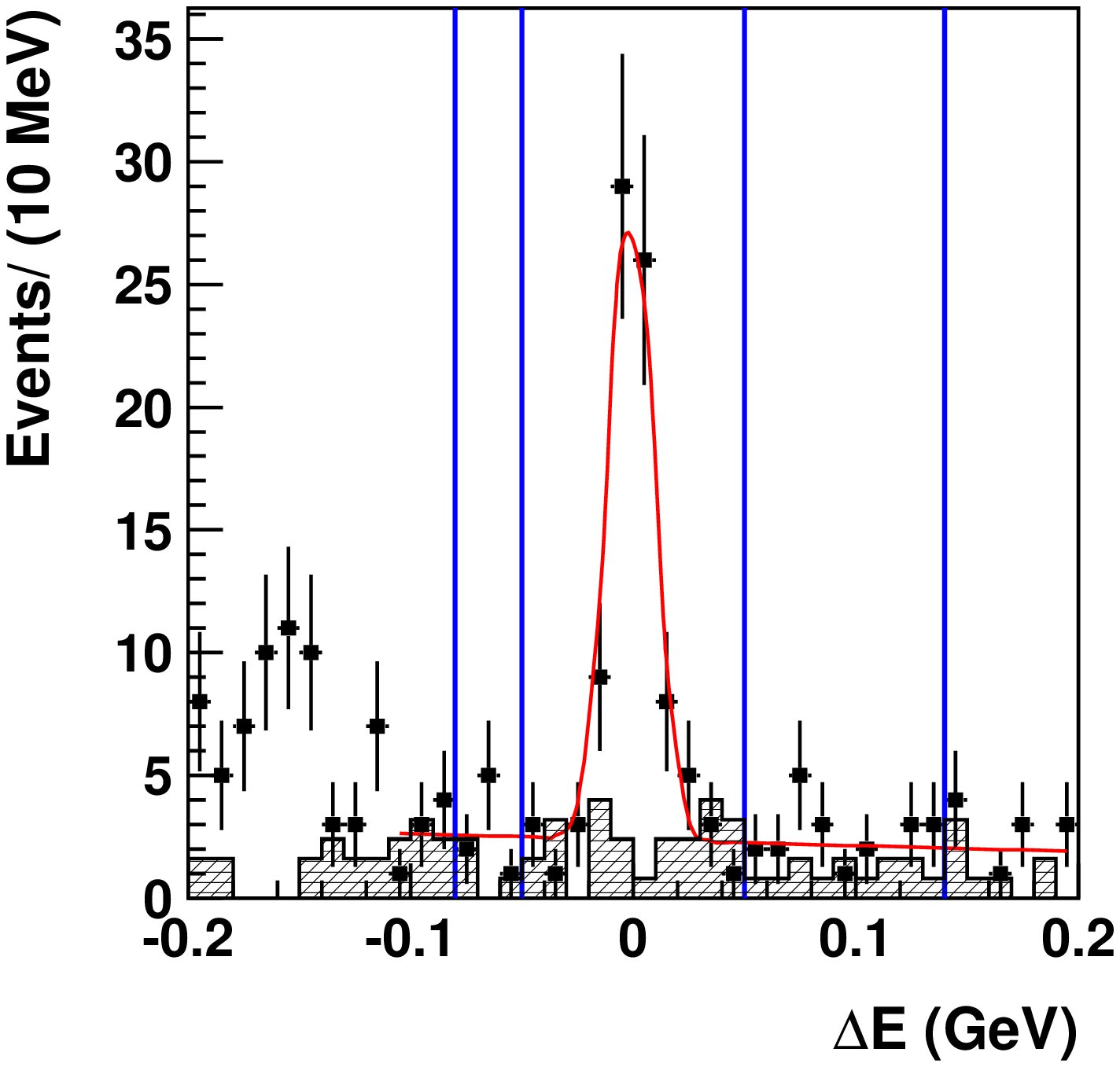}}
\resizebox{5.4cm}{!}{\includegraphics{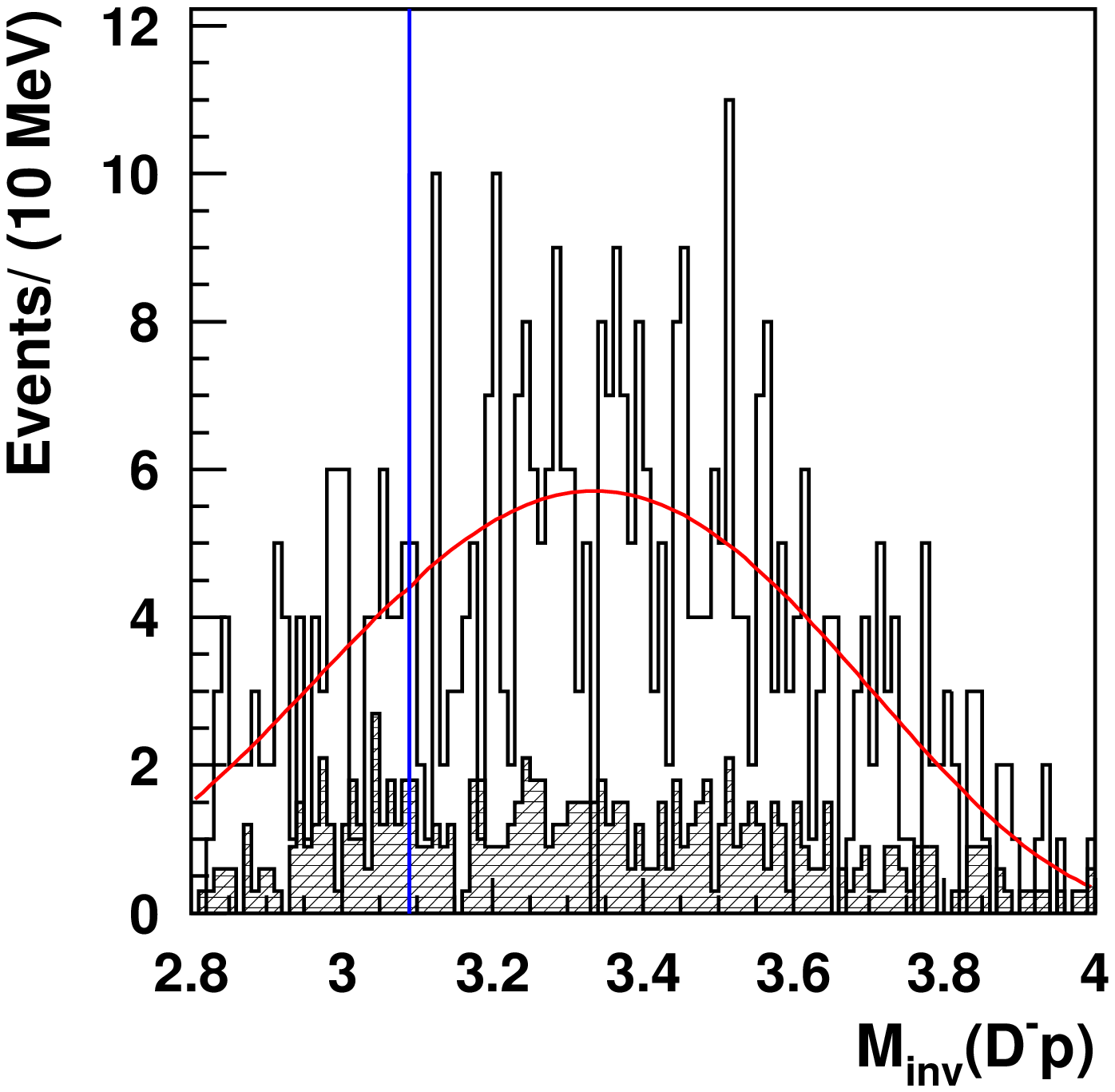}}
\resizebox{5.4cm}{!}{\includegraphics{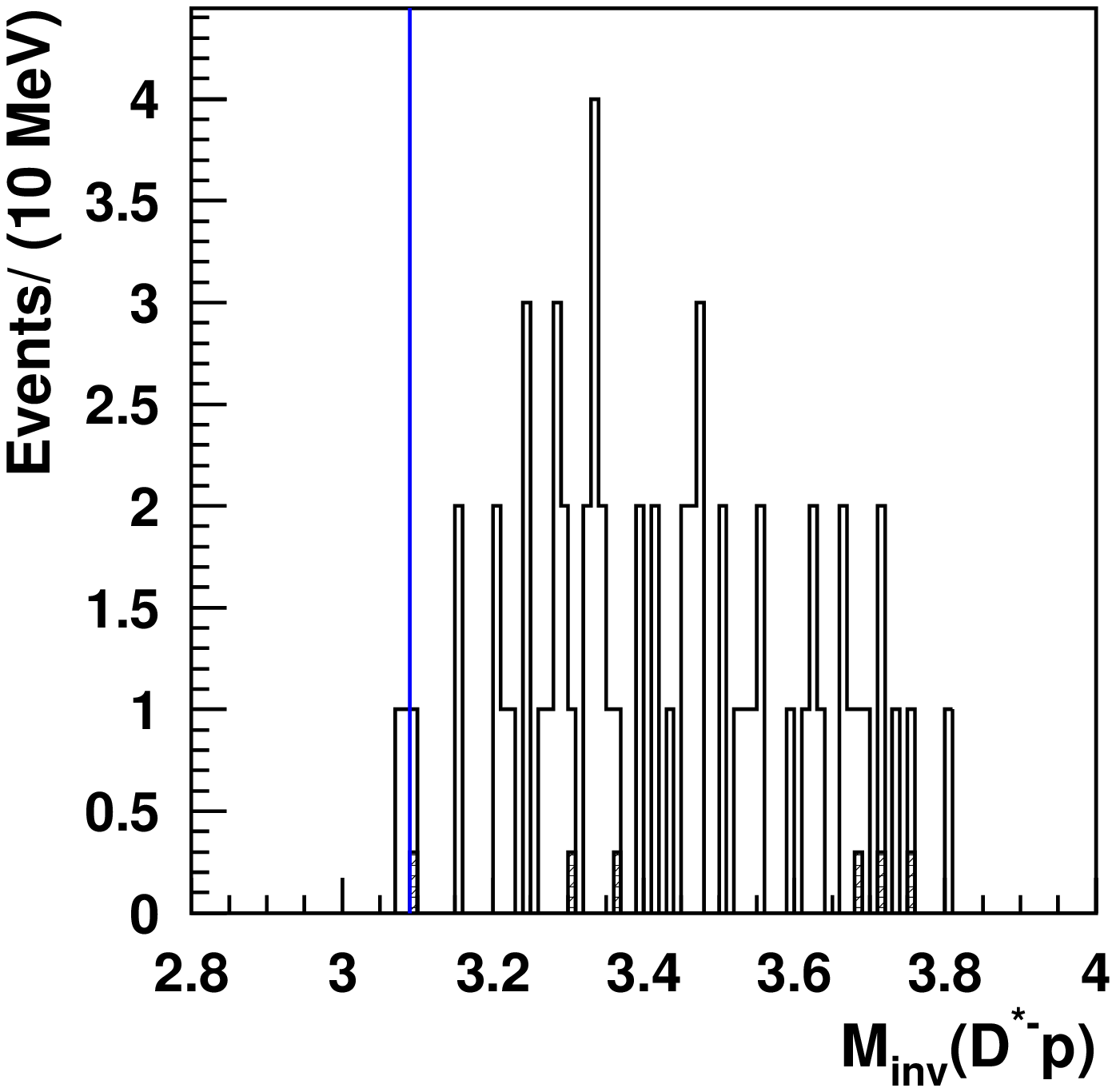}}
\resizebox{5.4cm}{!}{\includegraphics{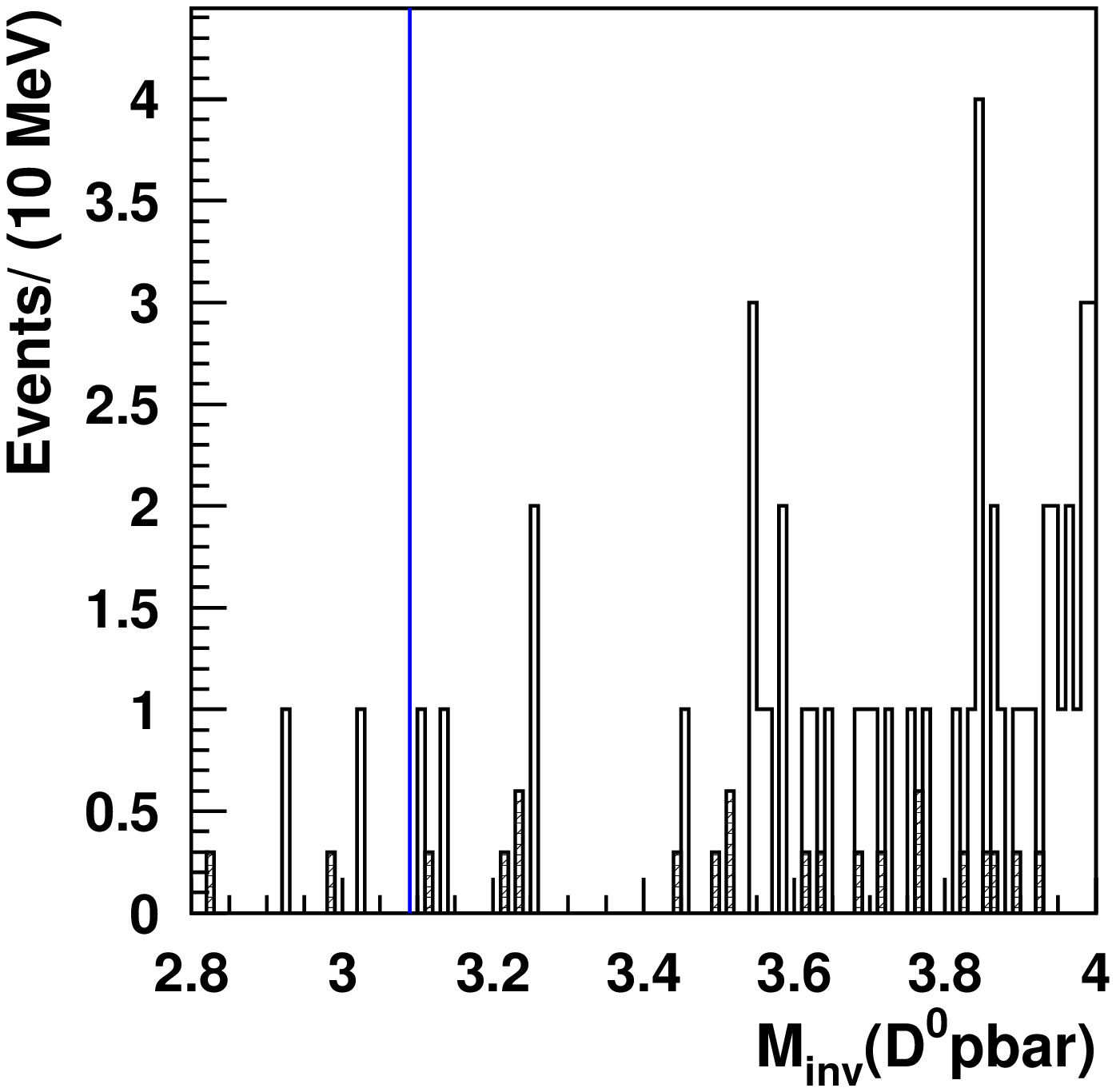}}
\caption{ 
  $\Delta E$ and $M(D^{(\ast)}p)$ distributions for
  $B^0 \to D^- p \bar{p} \pi^+$ (left),
  $B^0 \to D^{\ast -} p \bar{p} \pi^+$ (middle) and
  $\bar{B}^0 \to D^0 p \bar{p}$ (right) decays.
  The hatched histogram in $\de$ distributions corresponds to the $D$
  meson sidebands, while the hatched histogram in $M(D^{(\ast)}p)$
  distributions corresponds to the $\de$ sidebands (shown with
  vertical lines on the $\de$ plots). The vertical line in the
  $M(D^{(\ast)}p)$ distributions shows the H1 $\thc$ mass,
  $3.099\,\gev$.
}
\label{theta_c}
\end{center}
\end{figure}

The $\de$ and corresponding $M(D^{(\ast)}p)$ plots for the decays
$B^0 \to D^- p \bar{p} \pi^+$, 
$B^0 \to D^{\ast -} p \bar{p} \pi^+$ and 
$\bar{B}^0 \to D^0 p \bar{p}$
are shown in Fig.~\ref{theta_c}.
From the fit to $\de$ spectra the numbers of reconstructed $B$ decays
are $303\pm 21$, $60\pm 8$ and $66\pm 9$ for the three modes,
respectively. 
No signal of $\thc$ or $\thcst$ is found in the $M(D^{(\ast)}p)$ spectra. 
We set the following upper limits on the fractions of the final state proceeding via
$\thc$ and $\thcst$:
\begin{align*}
\frac{{\mathcal B}(B^0 \to \thc \bar{p} \pi^+) \times 
              {\mathcal B}(\thc \to D^{-} p)}
     {{\mathcal B}(B^0 \to D^- p \bar{p} \pi^+)} & < 1.2\%~(90\%~\text{CL}),
\\
\frac{{\mathcal B}(B^0 \to \thc \bar{p} \pi^+) \times 
              {\mathcal B}(\thc \to D^{\ast-} p)}
     {{\mathcal B}(B^0 \to D^{\ast -} p \bar{p} \pi^+)} & < 11\%~(90\%~\text{CL}),
\\
\frac{{\mathcal B}(B^0 \to \thcst \bar{p}) \times 
              {\mathcal B}(\thcst \to \bar{D^0} p)}
     {{\mathcal B}(\bar{B}^0 \to D^0 p \bar{p})} & < 5.9\%~(90\%~\text{CL}).
\end{align*}
We assume here that the charmed pentaquark mass is $3.099\,\gev$ 
and that the signal p.d.f.\ is determined by the detector resolution 
\mbox{($\sim 3.5\,\mev$)}~\cite{h1}. Our limits can be compared with the H1
claim that about 1\% of $D^{\ast +}$ mesons originate from $\thc$
decays. The branching fraction for $\thc$ decays to $D$ mesons 
is expected to be even larger because of the larger phase space.

\section{Acknowledgments}

We thank the KEKB group for the excellent operation of the
accelerator, the KEK Cryogenics group for the efficient operation
of the solenoid, and the KEK computer group and the NII for
valuable computing and Super-SINET network support.  We
acknowledge support from MEXT and JSPS (Japan); ARC and DEST
(Australia); NSFC (contract No.~10175071, China); DST (India); the
BK21 program of MOEHRD and the CHEP SRC program of KOSEF (Korea);
KBN (contract No.~2P03B 01324, Poland); MIST (Russia); MESS
(Slovenia); NSC and MOE (Taiwan); and DOE (USA).

\end{document}